\documentclass[a4paper,11pt]{article}
\usepackage{pos}

\usepackage{amsmath,amssymb}
\definecolor{red}{rgb}{1,0,0}

\graphicspath{{./figures/}}

\title{Moat Regimes in QCD and their Signatures\\ in Heavy-Ion Collisions}

\ShortTitle{Moat Regimes in QCD and their Signatures in Heavy-Ion Collisions}

\author*[a]{Fabian Rennecke}
\author[b]{Robert D.\ Pisarski}

\affiliation[a]{Institute for Theoretical Physics\\
Justus Liebig University Giessen\\
Heinrich-Buff-Ring 16, 35392 Giessen, Germany}

\affiliation[b]{Physics Department,\\
Brookhaven National Laboratory\\
Upton, NY 11973, USA}

\emailAdd{fabian.rennecke@theo.physik.uni-giessen.de}
\emailAdd{pisarski@bnl.gov}

\abstract{
Dense QCD matter can exhibit spatially modulated regimes. They can be characterized by particles with a moat spectrum, where the minimum of the energy is over a sphere at nonzero momentum. Such a moat regime can either be a precursor for the formation inhomogeneous condensates, or signal a quantum pion liquid.
We introduce the quantum pion liquid and discuss the underlying physics of the moat regime based on studies in low-energy models and preliminary results in QCD. 
Heavy-ion collisions at small beam energies have the potential to reveal the rich phase structure of QCD at low temperature and nonzero density. We show how moat regimes can be discovered through such collisions. Particle production is enhanced at the bottom of the moat, resulting in a peak at nonzero momentum, instead of zero, in the particle spectrum. Particle number correlations can even increase by several orders of magnitude at nonzero momentum in the moat regime.
}

\FullConference{%
	The International Conference on Critical Point and Onset of Deconfinement - CPOD2021\\
  	15 – 19 March 2021\\
 	Online - zoom
 }

\begin{document}
\maketitle

\section{Introduction}

QCD has a potentially rich phase structure at finite temperature and density \cite{Fukushima:2010bq}. It is of fundamental importance for the formation of matter around us and the composition of neutron stars. Experimentally, the phase diagram at different temperatures and densities is probed by beam energy scans in heavy-ion collisions \cite{Luo:2020pef}. Theoretically, it is notoriously hard to study strongly interacting matter at finite density, since a sign problem severely restricts the applicability of first-principles methods that rely on importance sampling of field configurations, such as lattice gauge theory. Thus, our current knowledge of the QCD phase diagram at finite density is largely based on model calculations. Viable alternatives to study QCD from first principles, which do not suffer from sign problems, are functional continuum methods, such as the functional renormalization group (FRG) \cite{Fu:2019hdw} and Dyson-Schwinger equations \cite{Fischer:2018sdj}. 

Here, we focus on regimes with periodic modulations of the spatial structure in the phase diagram. Prominent examples are inhomogeneous, or crystalline, phases, where quark- (in hadronic matter) or diquark- (in color-superconducting matter) condensates are spatially modulated \cite{Fukushima:2010bq,Buballa:2014tba}. Furthermore, there can be regimes where no condensation occurs, but correlation function still show spatial modulations. These regimes are called quantum pion liquids (Q$\pi$L) \cite{Pisarski:2020dnx, Pisarski:2021aoz, Tsvelik:2021ccp}. It turns out that a simple mean-field picture is insufficient here, as quantum fluctuations play a decisive role.
After briefly reviewing the basic physics, we show what signatures they can leave in heavy-ion collisions. 

\section{Moat regimes in the QCD phase diagram}

Model studies suggest that regimes with periodic spatial modulations can occur at high baryochemical potential $\mu_B$. In fact, phases with inhomogeneous chiral condensates are ubiquitous in lower-dimensional systems, e.g.\ \cite{Thies:2006ti}. A simple example is an $O(N_f)$ chiral density wave,
\begin{align}
\big\langle \vec{\phi}\, \big\rangle = \Delta \big( \cos(k_0\, z),\, \sin(k_0\, z),\dots,0 \big)^T\,,
\end{align} 
where the $O(N_f)$-vector condensate oscillates with wavenumber $k_0$ between two field components while it moves in one spatial ($z$) direction. In addition to chiral symmetry, such a condensate also breaks continuous translational invariance and, inevitably, rotational invariance. This leads to Goldstone bosons associated with chiral symmetry breaking if $N_f > 2$, and Goldstone bosons that result from the spontaneous breakdown of spatial symmetries. According to the Landau-Peierls theorem, fluctuations in transverse spatial direction around a one-dimensional inhomogeneous condensate lead to \emph{logarithmic} divergences \cite{Landau:1980mil}. These fluctuations stem from the Goldstone bosons associated to spatial symmetry breaking. They disorder the condensate at large distances, leading to only a quasi-long range order in the system.

However, also the Goldstone bosons of chiral symmetry breaking, which in our simple example of a chiral density wave arise whenever $N_f > 2$, lead to fluctuations in transverse isospin direction that disorder the inhomogeneous condensate. To see this, we note that these fields are favored to propagate with a momentum related to the wavenumber $k_0$ of the spatial oscillation. Their static propagator then is
\begin{align}
G_{\rm GB}^{-1}({\mathbf p}^2) = W \big( {\mathbf p}^2 - k_0^2 \big)^2\,.
\end{align}
This is true for a general form of the effective potential \cite{Pisarski:2020dnx}. These Goldstone bosons are massless at the wavenumber ${\mathbf p}^2 = k_0^2$, but through a double pole at this point. Thus, any tadpole diagram at finite $T$ in the inhomogeneous phase involving these transverse modes gets a contribution proportional to
\begin{align}
T \int\!  \frac{d^3{\mathbf p}}{(2\pi)^3} G_{\rm GB} \sim \frac{T}{W} \int_{|{\mathbf p}|\sim k_0}\frac{d|{\mathbf p}|}{(|{\mathbf p}|-k_0)^2}\,. 
\end{align}
This leads to a severe \emph{linear} infrared divergence. Our simple example illustrates that fluctuations in transverse isospin direction destabilize the system so that the inhomogeneous order is destroyed. As opposed to the logarithmic divergence resulting from transverse spatial fluctuations, there is not even quasi-long range order. In addition to the perturbative analysis, this has also been established more generally in the limit of large number of flavors, $N_f\rightarrow\infty$, in \cite{Pisarski:2020dnx}. At zero temperature, there still is a logarithmic divergence \cite{Pisarski:2021aoz}. These analyses show that, instead of an inhomogeneous phase, there is a stable symmetric phase where a mass gap for the transverse modes is generated dynamically. But as opposed to an ordinary symmetric phase, the underlying spatial modulation still leaves imprints in the correlations of the system. Rather than a double poles at $k_0$, the transverse propagator has complex poles $m_r + i m_i$, which lead to an oscillation on top the exponential fall-off of the two-point function at large separation,
\begin{align}\label{eq:2pt}
\langle \phi(x) \phi(0) \rangle\big|_{x\rightarrow \infty}\,\sim\; e^{- m_r\, x} \cos(m_i\, x)\,.
\end{align}
In analogy to quantum spin liquids, this regime is called Q$\pi$L. While spatial rotational symmetry is intact in this phase, boost symmetry is broken.

\begin{figure}[t]
\centering
\includegraphics[width=.48\columnwidth]{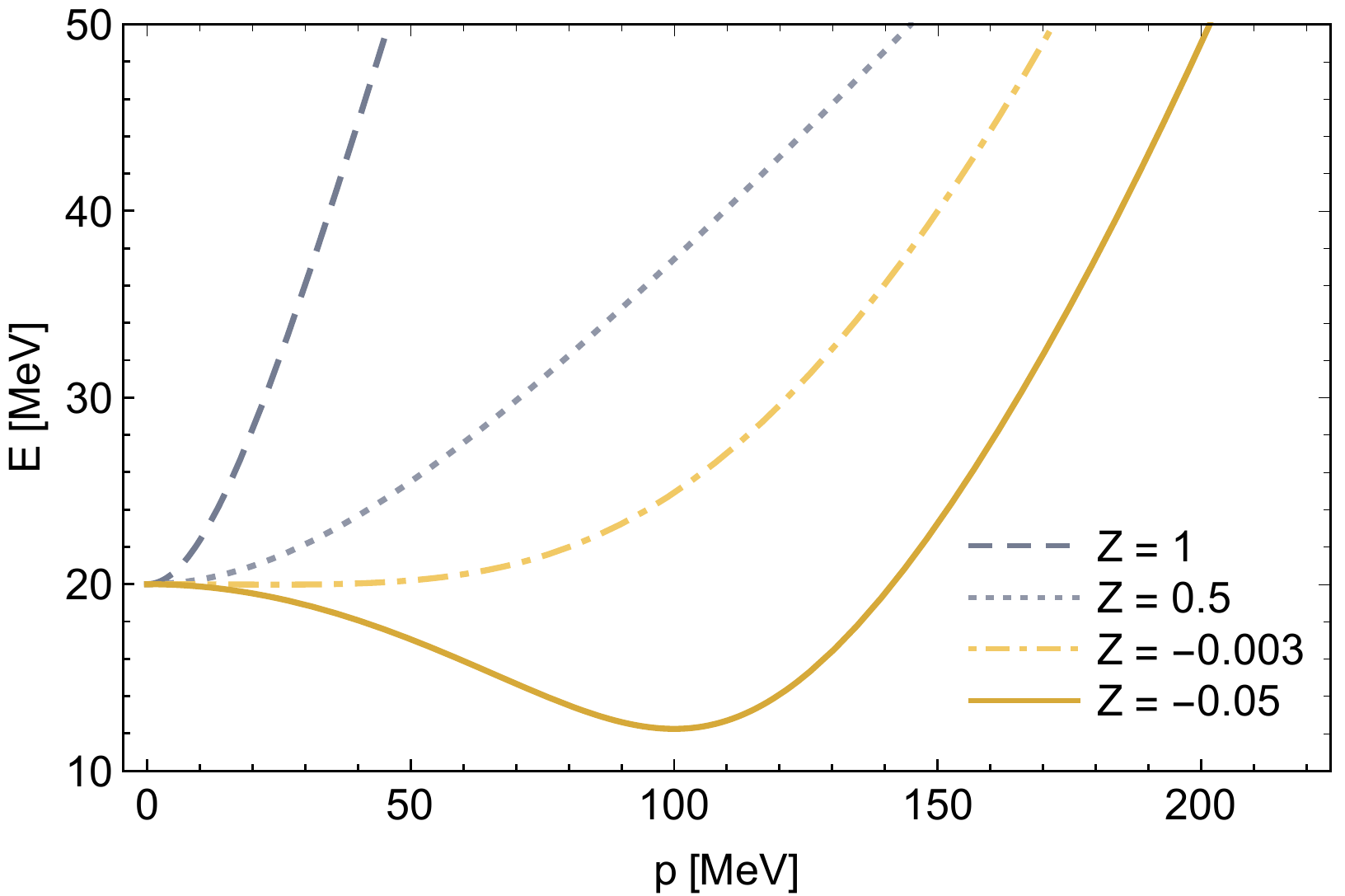}
\hfill
\includegraphics[width=.5\columnwidth]{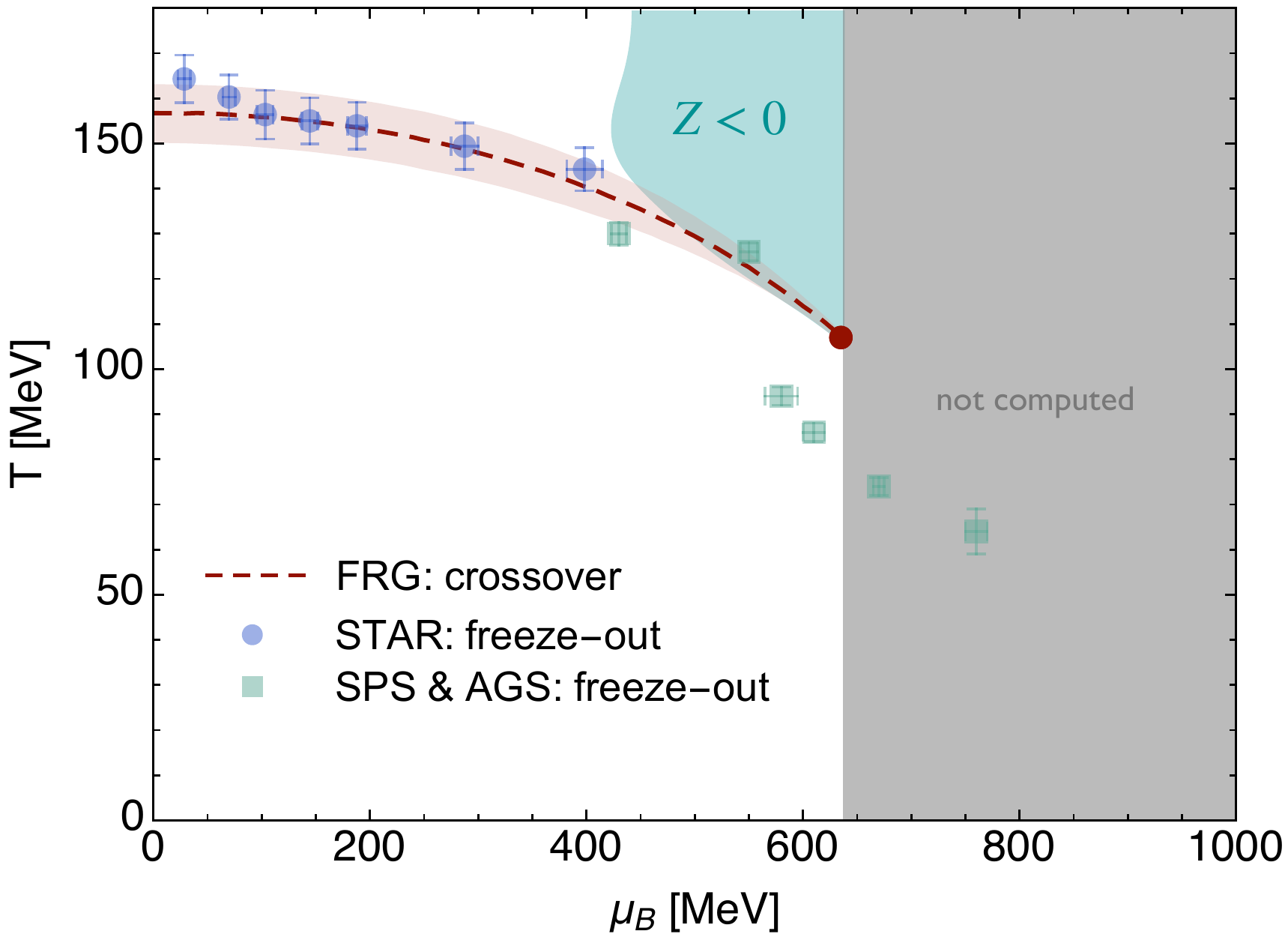}
\caption{\emph{Left:} Dispersion relations in the normal ($Z>0$, gray lines) and the moat ($Z<0$, yellow lines) regime as functions of the spatial momentum.
\emph{Right:} QCD phase diagram computed in with the FRG in \cite{Fu:2019hdw} (the dashed red line denotes the chiral crossover, the shaded red region the width of the crossover and the red dot the location of the critical endpoint), together with heavy-ion freeze-out points from \cite{STAR:2017sal} and \cite{Andronic:2017pug}. The shaded green area ($Z<0$) shows where indications for a moat regime have been found.}
\label{fig:disp}
\end{figure}

In rotationally invariant but boost variant systems, the energy dispersion $E_\phi$ of the field $\phi$ at small momentum can be written as
\begin{align}\label{eq:moat}
E_\phi({\mathbf p}^2) = \sqrt{ Z\, {\mathbf p}^2 + W ({\mathbf p}^2)^2 + m_{\rm eff}^2 }\,,
\end{align}
with some coefficients $Z$ and $W > 0$ that depend on the temperature and the chemical potential. In a dense medium one can have $Z<0$. In this case the energy in Eq.~(\ref{eq:moat}) is minimal at a nonzero momentum,
\begin{align}
{\mathbf p}_{\rm min}^2 = -\frac{Z}{2 W} \equiv k_0^2\,,
\end{align}
and looks like a moat, as shown in the left plot of Fig.~\ref{fig:disp}. We therefore call the regime in the phase diagram with $Z<0$ a \emph{moat regime}. As already mentioned, the fact that the particle is favored to have a finite momentum $|{\mathbf p}| = k_0$ is directly related to an underlying spatial modulation with wavenumber $k_0$. This is in particular true in the Q$\pi$L, where the large-$N_f$ analysis shows that the wave number in Eq.~(\ref{eq:2pt}) indeed is $m_i = k_0$ for large, negative $Z$ \cite{Pisarski:2020dnx}. $E_\phi({\mathbf p}^2) > 0$ always holds for all momenta, since a mass gap is dynamically generated in the  Q$\pi$L.

Inhomogeneous phases occur whenever $E_\phi(k_0^2) = 0$, since then condensation of field modes with momentum $k_0$ is favored. Thus, a moat regime in the QCD phase diagram can either be sign of a Q$\pi$L or a precursor for an inhomogeneous phase. First indications of the existence of a moat regime have been found in first-principles studies of the phase diagram with the FRG in \cite{Fu:2019hdw}. The result is shown in the right plot of Fig.~\ref{fig:disp}. The shaded green area with $Z<0$ could either be a stable Q$\pi$L or indicate that, for example, an inhomogeneous chiral or diquark condensate appears at lower $T$ and larger $\mu_B$. We emphasize however, that the fluctuation analysis presented above strongly favors a Q$\pi$L over any inhomogeneous phase with massless transverse modes. Further studies are required to sort this out.

\section{Signatures of moat regimes in heavy-ion collisions}

Since the energy of particles in a moat regime is minimal at nonzero momentum, their production in heavy-ion collisions should be enhanced at a spatial momentum which is directly related to the wavenumber $k_0$ \cite{Pisarski:2020gkx}. We have shown in \cite{Pisarski:2021qof} that transverse momentum spectra and correlations can indeed show characteristic peaks at nonzero momentum.

Particles in the expanding medium of a heavy-ion collision freeze out at a certain $T$ and $\mu_B$, and then stream to the detector \cite{Andronic:2017pug}. This defines the freeze-out surface $\Sigma$, a $3d$ hypersurface of the $4d$ spacetime-volume of the system. We therefore need to describe the particles and their correlations on $\Sigma$. Since the moat regime arises through strong interaction effects in the dense medium, $\Sigma$ is not the kinetic freeze-out surface in our case, as the system is dilute there. However, it can be identified with the chemical freeze-out surface, which is close to the chiral phase transition and can still have significant interactions. In general, we assume $\Sigma$ to be a hypersurface at fixed $T$ and $\mu_B$ in the moat regime. On this surface, particles move with the fluid velocity $u^\mu(x)$, so that their energy $\breve p_0 = u^\mu p_\mu$ and spatial momentum $\breve{\mathbf p}^2\, = (u^\mu u^\nu - g^{\mu\nu}) p_\mu p_\nu$ ($g^{\mu\nu}$ is the Minkowski metric) are boosted accordingly.

Momentum spectra of particles on $\Sigma$ are usually described by the Cooper-Frye formula \cite{Cooper:1974mv}. This formula is not applicable in our case as it assumes free particles, whereas we consider an interacting medium. As shown in \cite{Pisarski:2021qof}, the Cooper-Frye formula can be generalized by using the Wigner function $F_\phi(p)$ in thermal equilibrium,
\begin{align}\label{eq:wigner}
F_\phi(p) = 2\pi\, \rho_\phi(p^0,\mathbf{p})\, f(p_0)\, ,
\end{align}
where $\rho_\phi$ is the spectral function of $\phi$ and $f(p_0)$ is the thermal single-particle distribution function, so, e.g., the Bose-Einstein distribution for bosons. $F_\phi(p)$ is the probability of finding a particle $\phi$ with four-momentum $p$. The differential particle spectrum is then obtained by integrating the resulting particle current density over $\Sigma$,
\begin{align}\label{eq:spec}
\frac{d^3 N_\phi}{d \mathbf{p}^3} 
=\frac{2}{(2\pi)^3} \int_\Sigma\, d\Sigma_\mu \int\! \frac{d p^0}{2\pi}\,p^\mu\, \Theta(\breve p^0)\, F_\phi(\breve p)\,.
\end{align}
The Heaviside-function $\Theta$ enforces positive energy. Thus, given the spectral function, the surface $\Sigma$ and the fluid velocity $u^\mu$, the particle spectrum can be computed from Eq.~(\ref{eq:spec}). 

To capture the basic properties of the particle spectrum, it is sufficient to use simple models. For the spectral function, we us a quasi-particle low-energy model of free bosons in a moat regime, which yields the spectral function $\rho_\phi(\breve p^0 , \breve{\mathbf p}^2) = \rm{sign}(\breve p^0)\, \delta\big[(\breve p^0)^2-E_\phi^2(\breve{\mathbf p}^2)\big]$.
$E_\phi$ is the moat dispersion given in Eq.~(\ref{eq:moat}). To get the fluid velocity and the freeze-out surface, we assume that the radial fluid velocity is proportional to the radial distance and that $\Sigma$ is boost invariant. This can be described by a blast-wave model, fitted to data at a beam energy of $\sqrt{s} = 5$~GeV \cite{Zhang:2016tbf}.  The corresponding chemical potential is $\mu_B = 536$~MeV and therefore consistent with where the moat regime is expected in the right plot of Fig.~\ref{fig:disp}. Thermodynamic properties of the system are assumed to be described by a hadron resonance gas \cite{Andronic:2017pug}.

\begin{figure}[t]
\centering
\includegraphics[width=.55\columnwidth]{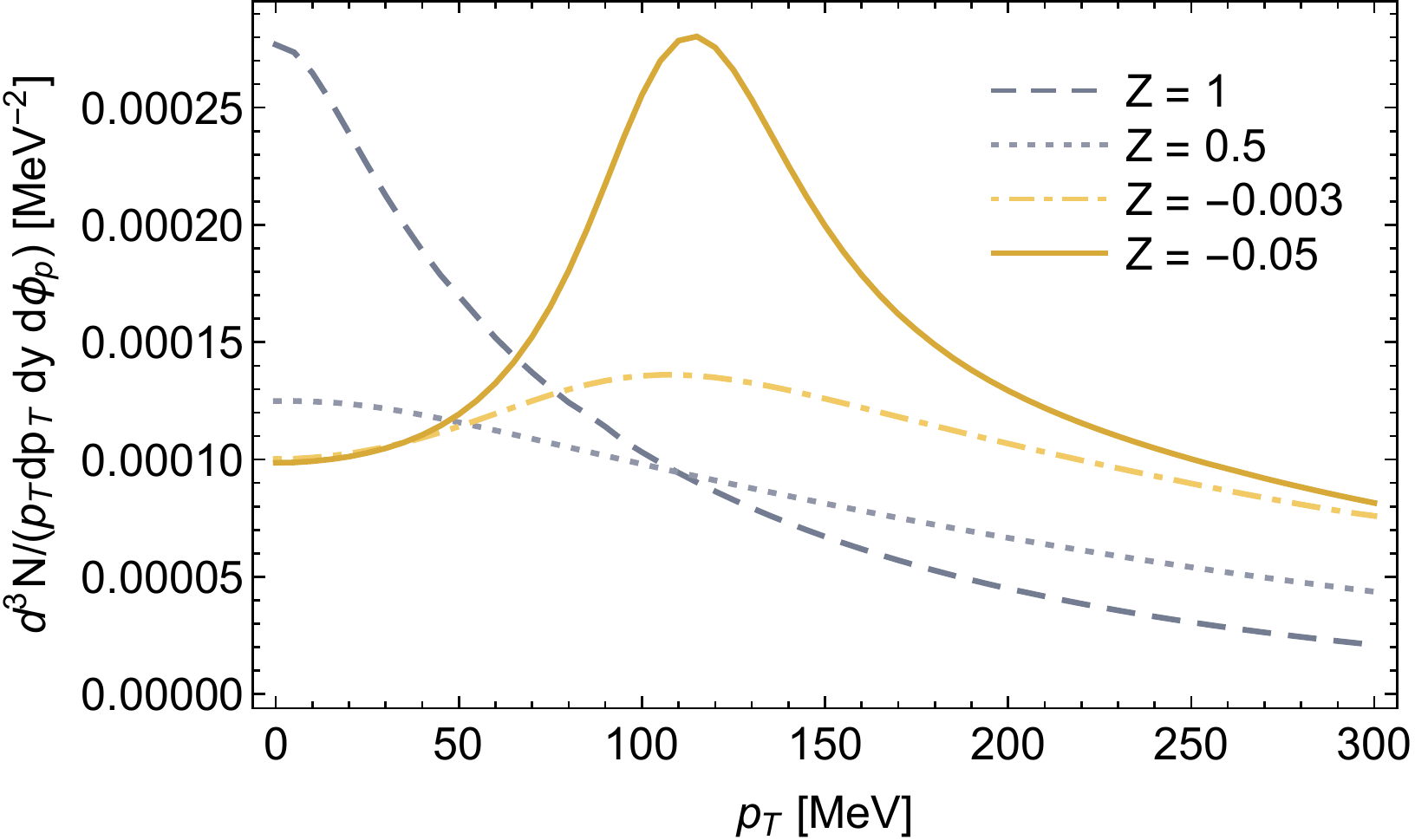}
\caption{Comparison between the transverse momentum spectrum in the normal ($Z>0$, gray lines)
  and the moat ($Z<0$, yellow lines) regime.}
\label{fig:spec}
\end{figure}

The resulting transverse momentum spectrum is shown in Fig.~\ref{fig:spec} for various values of $Z$ and fixed $W>0$ (the same parameters as in Fig.~\ref{fig:disp}, left). As anticipated, the spectrum indeed shows a pronounced peak at nonzero momentum in the moat regime. The location of this peak is directly related to the boosted wave number $k_0$. Since these spectra in the normal phase with $Z>0$ always peak at zero momentum, this is a clear signature of a moat regime.

In addition to the spectrum, one can also consider particle correlations on $\Sigma$. Since the moat regime is in a disordered phase, long-range correlations are absent. This greatly simplifies the problem, since multi-particle correlations can be computed directly from correlations of Eq.~(\ref{eq:spec}) \cite{Pisarski:2021qof}:
\begin{align}\label{eq:corgen}
\Bigg\langle \prod_{i} \frac{d^3 N_\phi}{d \mathbf{p}_i^3}  \Bigg\rangle &= \Bigg[\prod_i \frac{2}{(2\pi)^3} \int\! d\Sigma^\mu_i \int\! \frac{dp_i^0}{2\pi}\,(p_i)_\mu\, \Theta(\breve p^0_i)\Bigg]
\bigg\langle\prod_i  F_\phi(\breve p_i) \bigg\rangle\,.
\end{align}
Correlations of particle numbers are generated through fluctuations of the Wigner function $F_\phi$. Assuming that that these fluctuations are thermodynamic in nature, so that $\kappa_i^\mu(x) = \big( T(x),\,\mu_B(x),\, u^\mu(x) \big)_i$ fluctuates, $\kappa = \bar\kappa + \delta\kappa$, fluctuations in $F_\phi$ are generated by fluctuations in $\kappa$. The connected two-point function, for example, then becomes $\big\langle F_\phi\,F_\phi  \big\rangle_c = \frac{\partial F_\phi}{\partial\kappa_i^\mu}  \frac{\partial F_\phi}{\partial\kappa_j^\nu}\bigg|_{\bar\kappa} \big\langle \delta\kappa_i^\mu \delta\kappa_j^\nu \big\rangle +\mathcal{O}\big(\delta\kappa^3\big)$.
The average $\langle \cdot \rangle_c$ can be computed from the generating functional of thermodynamic correlations on $\Sigma$,
\begin{align}\label{eq:genfunc}
  {\rm e}^{W[J]} = \int\!\!\!\mathcal{D}\kappa \exp\! \int\!\!\!
  d\Sigma_\mu  \left[\Delta s^\mu(x) + \ J(x)_{i\nu}\, \hat{v}^\mu\delta\kappa^\nu_i(x)  \right] \,,
\end{align}
by taking derivatives with respect to the source $J$ at $J=0$. $\hat v$ is a four-vector normal to $\Sigma$. The configurations are weighted by the change in the entropy current density due to fluctuations in each fluid cell, $\Delta s^\mu$. It is given the difference between the entropy current density of the subsystem away from equilibrium, specified by $\bar\kappa + \delta\kappa$, and that in equilibrium at $\bar\kappa$.

\begin{figure}[t]
\centering
\includegraphics[width=.49\columnwidth]{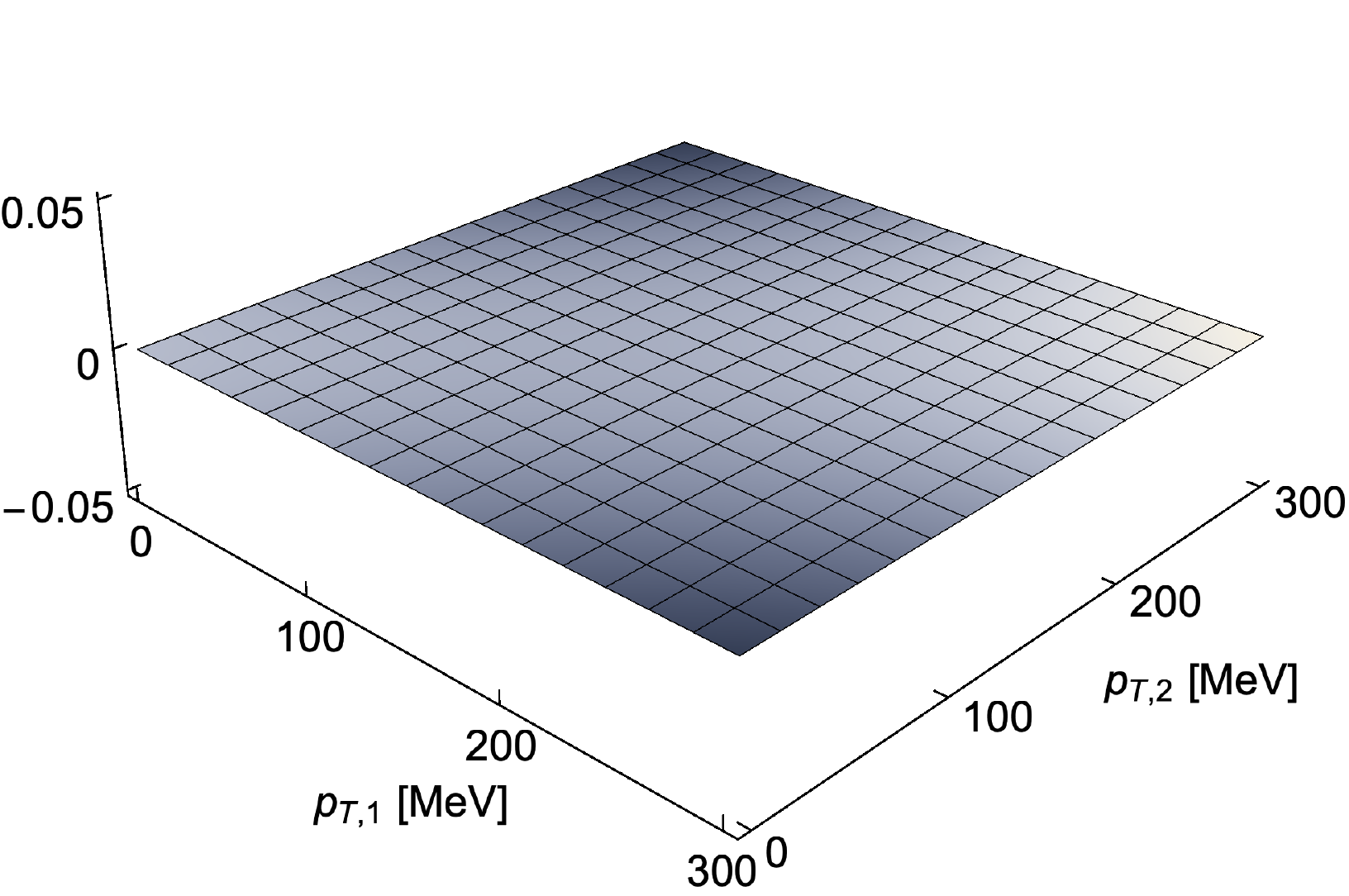}
\hfill
\includegraphics[width=.49\columnwidth]{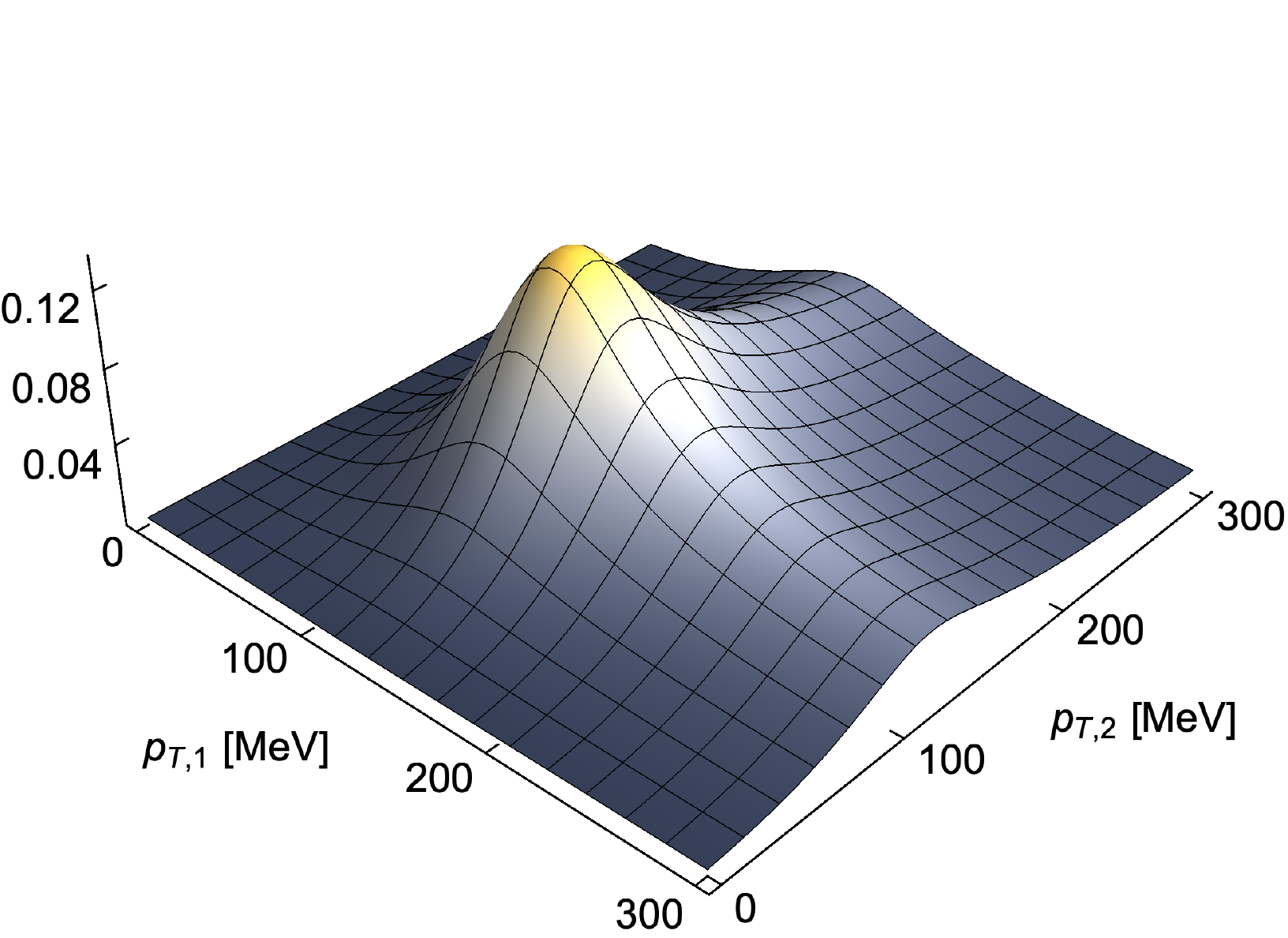}
\caption{Connected two-particle correlation normalized with the spectra, $\Delta n_{12} = \big\langle \big(\frac{d^3N}{d\bold{p}^3}\big)^2 \big\rangle_c\, \big/ \big\langle \frac{d^3N}{d\bold{p}^3} \big\rangle^2$, as a function of the transverse momenta of the two particles.
\emph{Left:} Normal phase.
\emph{Right:} Moat regime.}
\label{fig:corr}
\end{figure}

For small fluctuations in an ideal fluid, this has been worked-out explicitly in \cite{Pisarski:2021qof}. The resulting connected two-particle $p_T$-correlation is shown in Fig.~\ref{fig:corr}. The left plot shows the result in the normal phase and the result in the moat regime is shown in the right plot. While the two-point function is essentially flat in the normal phase, it has a pronounced peak and ridges at nonzero momenta in the moat regime. The enhancement at the peak is about two orders of magnitude and the peak location is again related to the boosted wavenumber $k_0$.

\section{Conclusions}

We argued that an essential feature of regimes with spatial modulations, which can occur in the QCD phase diagram at large density, is a moat energy spectrum. It signals either a Q$\pi$L or is a precursor for inhomogeneous condensation. Since the energy of particles is minimal at the bottom of the moat, particle production is enhanced at nonzero momentum in a moat regime. This manifests itself in pronounced peaks at nonzero momentum in $p_T$-resolved particle spectra and correlations. Thus, if the matter created in low-energy heavy-ion collisions traverses a moat regime, these peaks, which are expected to occur at relatively low $p_T$, can be clear signatures of regimes with spatial modulations in QCD.

\acknowledgments{
This research was supported by the U.S.\ Department of Energy under contract DE-SC0012704 and by B.N.L.\ under the Lab Directed Research and Development program 18-036.
}

\vspace{-0.2cm}
\bibliographystyle{JHEP}
\bibliography{moat}

\end{document}